\documentclass[11pt,letterpaper]{article}
\usepackage[T1]{fontenc}
\usepackage[utf8]{inputenc}
\usepackage{lmodern}
\usepackage[margin=1in]{geometry}
\usepackage{graphicx}
\usepackage{amsmath}
\usepackage{microtype}
\usepackage{placeins}
\usepackage{caption}
\usepackage[super,sort&compress,comma]{natbib}
\usepackage[breaklinks,hidelinks]{hyperref}
\newcommand{\protocolstep}[2]{%
  \par\addvspace{0.35\baselineskip}\noindent
  \hangindent=4.3em\hangafter=1
  \makebox[4.3em][l]{\textbf{#1}}#2\par}
\newcommand{\doi}[1]{\href{https://doi.org/#1}{doi:\nolinkurl{#1}}}
\title{A Workflow for Lipid Nanoparticle (LNP) Formulation Optimization Using Designed Mixture-Process Experiments and Self-Validated Ensemble Models (SVEM)}
\author{Andrew T. Karl\textsuperscript{1}, Sean Essex\textsuperscript{2},\\
James Wisnowski\textsuperscript{1}, Heath Rushing\textsuperscript{1}\\[0.5em]
\small\textsuperscript{1}Adsurgo, LLC; San Antonio, Texas, USA\\
\small\textsuperscript{2}Arcturus Therapeutics; San Diego, California, USA}
\date{}
\hypersetup{pdftitle={A Workflow for Lipid Nanoparticle Formulation Optimization},pdfauthor={Andrew T. Karl, Sean Essex, James Wisnowski, Heath Rushing}}
\begin{document}
\maketitle
\section*{Summary}
This protocol provides an approach to formulation optimization over mixture, continuous, and categorical study factors that minimizes subjective choices in the experimental design construction. For the analysis phase, an effective and easy-to-use model-fitting procedure is employed.

\begin{abstract}
We present a Quality by Design (QbD) styled approach for optimizing lipid nanoparticle (LNP) formulations, aiming to offer scientists an accessible workflow. The inherent restriction in these studies, where the molar ratios of ionizable, helper, and PEG lipids must add up to 100\%, requires specialized design and analysis methods to accommodate this mixture constraint. Focusing on lipid and process factors that are commonly used in LNP design optimization, we provide steps that avoid many of the difficulties that traditionally arise in the design and analysis of mixture-process experiments by employing space-filling designs and utilizing the recently developed statistical framework of self-validated ensemble models (SVEM). In addition to producing candidate optimal formulations, the workflow also builds graphical summaries of the fitted statistical models that simplify the interpretation of the results. The newly identified candidate formulations are assessed with confirmation runs, which can optionally be conducted in the context of a more comprehensive second-phase study.
\end{abstract}

\section*{Publication Notice}
\begingroup\small
The definitive article and video appeared in the \textit{Journal of Visualized Experiments} (JoVE), issue 198, e65200 (2023), \doi{10.3791/65200}. For this \href{https://arxiv.org/abs/2212.11264}{author edition}, Astra, an AI assistant, assisted with conversion to \LaTeX{} and minor clarifications and repairs; Claude Opus and Fable assisted with proofreading. The edition also includes pointers to later work on whole-model testing and SVEMnet\cite{karl2024wholemodel,karl2026svemnet}.

From version~4.0.0, SVEMnet retains its SVEM modeling and optimization functionality and adds other learners, such as Gaussian-process and extremely randomized-tree models, each able to nominate a candidate recipe alongside SVEM for physical confirmation. Expanded worked examples, including \textit{Choosing and confirming follow-up recipes in a 23-run lipid screen}, and further developments will be documented in the package vignettes. Check the released version and available vignettes at \url{https://CRAN.R-project.org/package=SVEMnet}.
\par\endgroup

\FloatBarrier
\section*{Introduction}

Lipid nanoparticle (LNP) formulations for \textit{in vivo} gene delivery systems generally involve four constituent lipids from the categories of ionizable, helper, and PEG lipids\cite{ref01,ref02,ref03}. Whether these lipids are being studied alone or simultaneously with other non-mixture factors, experiments for these formulations require ``mixture'' designs because -- given a candidate formulation -- increasing or decreasing the ratio of any one of the lipids necessarily leads to a corresponding decrease or increase in the sum of the ratios of the other three lipids.

For illustration, we will suppose that we are optimizing an LNP formulation that currently uses a set recipe that will be treated as the benchmark. The goal is to maximize the potency of the LNP while secondarily aiming to minimize the average particle size. The study factors that are varied in the experiment are the molar ratios of the four constituent lipids (ionizable, cholesterol, DOPE, PEG), the N:P ratio, the flow rate, and the ionizable lipid type. The ionizable and helper lipids (including cholesterol) are allowed to vary over a wider range of molar ratio, 10-60\%, than PEG, which will be varied from 1-5\% in this illustration. The benchmark formulation recipe and the ranges of the other factors and their rounding granularity are specified in \textbf{Supplementary File 1}. For this example, the scientists are able to perform 23 runs (unique batches of particles) in a single day and would like to use that as their sample size if it meets the minimum requirements. Simulated results for this experiment are provided in \textbf{Supplementary File 2} and \textbf{Supplementary File 3.}

Rampado and Peer\cite{ref04} provide a recent review paper on the topic of designed experiments for the optimization of nanoparticle-based drug delivery systems. Kauffman et al.\cite{ref05} consider LNP optimization studies using fractional factorial and definitive screening designs\cite{ref06}; however, these types of designs cannot accommodate a mixture constraint without resorting to the use of inefficient ``slack variables''\cite{ref07} and are not typically used when mixture factors are present\cite{ref07,ref08}. Instead, ``optimal designs'' capable of incorporating a mixture constraint are traditionally used for mixture-process experiments\cite{ref09}. These designs target a user-specified function of the study factors and are only optimal (in one of a number of possible senses) if this function captures the true relationship between the study factors and responses. Note that there is a distinction in the text between ``optimal designs'' and ``optimal formulation candidates'', with the latter referring to the best formulations identified by a statistical model. Optimal designs come with three main disadvantages for mixture-process experiments. First, if the scientist fails to anticipate an interaction of the study factors when specifying the target model, then the resulting model will be biased and can produce inferior candidate formulations. Second, optimal designs place most of the runs on the exterior boundary of the factor space. In LNP studies, this can lead to a large number of lost runs if the particles do not form correctly at any extremes of the lipid or process settings. Third, scientists often prefer to have experimental runs on the interior of the factor space to gain a model-independent sense of the response surface and to observe the process directly in previously unexplored regions of the factor space.

An alternative design principle is to target an approximate uniform coverage of the (mixture-constrained) factor space with a space-filling design\cite{ref10}. These designs sacrifice some experimental efficiency relative to optimal designs\cite{ref09} (assuming the entire factor space leads to valid formulations) but present several benefits in a trade-off that are useful in this application. The space-filling design does not make any \textit{a priori} assumptions about the structure of the response surface; this gives it the flexibility to capture unanticipated relationships between the study factors. This also streamlines the design generation because it does not require making decisions about which regression terms to add or remove as the desired run size is adjusted. When some design points (recipes) lead to failed formulations, space-filling designs make it possible to model the failure boundary over the study factors while also supporting statistical models for the study responses over the successful factor combinations. Finally, the interior coverage of the factor space allows for model-independent graphical exploration of the response surface.

\begin{figure}[htbp]
\centering
\includegraphics[width=\linewidth,height=0.67\textheight,keepaspectratio]{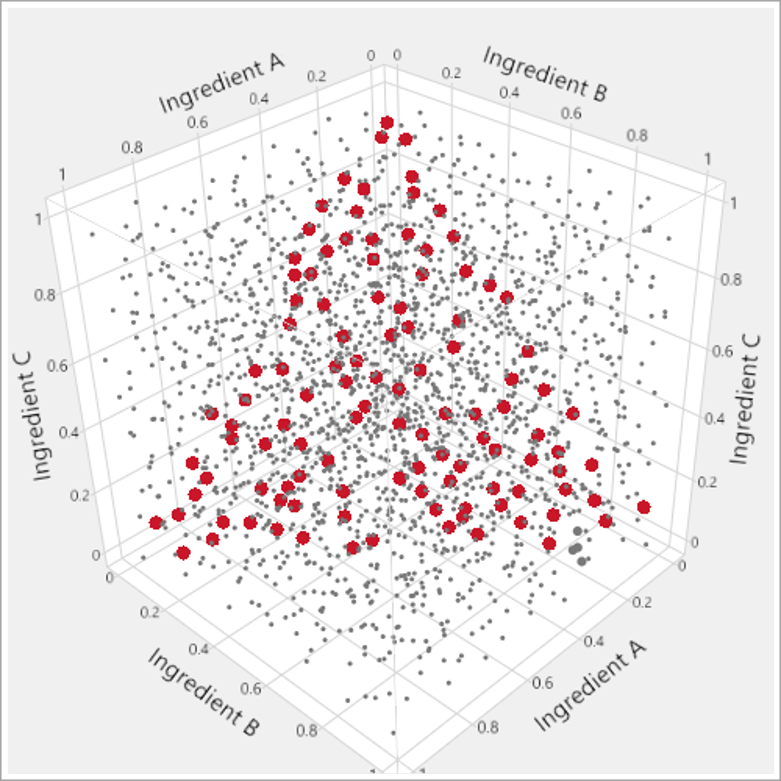}
\caption{Triangular factor regions. In the space-filling plot within the cube, the small grey dots represent formulations that are inconsistent with the mixture constraint. The larger red points lie on a triangle inscribed within the cube and represent formulations for which the mixture constraint is satisfied.}
\label{fig:1}
\end{figure}

To visualize the mixture factor subspace of a mixture-process experiment, specialized triangular ``ternary plots'' are used. \textbf{Figure~\ref{fig:1}} motivates this usage: in the cube of points where three ingredients are each allowed to range from 0 to 1, the points that satisfy a constraint that the sum of the ingredients equals 1 are highlighted in red. The mixture constraint on the three ingredients reduces the feasible factor space to a triangle. In LNP applications with four mixture ingredients, we produce six different ternary plots to represent the factor space by plotting two lipids at a time against an ``Others'' axis that represents the sum of the other lipids.

In addition to the lipid mixture factors, there are often one or more continuous process factors such as N:P ratio, buffer concentration, or flow rate. Categorical factors may be present, such as ionizable lipid type, helper lipid type, or buffer type. The goal is to find a formulation (a mixture of lipids and settings for process factors) that maximizes some measure of potency and/or improves physicochemical characteristics such as minimizing particle size and PDI (polydispersity index), maximizing percent encapsulation, and minimizing side effects -- such as body weight loss -- in \textit{in vivo} studies. Even when starting from a reasonable benchmark recipe, there may be interest in re-optimizing given a change in the payload or when considering changes in the process factors or lipid types.

Cornell\cite{ref07} provides a definitive text on the statistical aspects of mixture and mixture-process experiments, with Myers, Montgomery, and Anderson-Cook\cite{ref09} providing an excellent summary of the most relevant mixture design and analysis topics for optimization. However, these works can overload scientists with statistical details and with specialized terminology. We feel that modern software for the design and analysis of experiments provides a robust solution that will sufficiently support most LNP optimization problems without having to appeal to the relevant theory. While more complicated or high-priority studies will still benefit from collaboration with a statistician and may employ optimal rather than space-filling designs, our goal is to improve the comfort level of scientists and to encourage optimization of LNP formulations without appealing to inefficient one-factor-at-a-time (OFAT) testing\cite{ref11} or simply settling for the first formulation that satisfies specifications.

In this article, a workflow is presented that utilizes statistical software to optimize a generic LNP formulation problem, addressing design and analysis issues in the order that they will be encountered. In fact, the method will work for general optimization problems and is not restricted to LNPs. Along the way, several common questions that arise are addressed and recommendations are provided that are grounded in experience and in simulation results\cite{ref12}. The recently developed framework of self-validated ensemble models (SVEM)\cite{ref13} has greatly improved the otherwise fragile approach to analyzing results from mixture-process experiments, and we use this approach to provide a simplified strategy for formulation optimization. While the workflow is constructed in a general manner that could be followed using other software packages, JMP 17 Pro offers SVEM along with the graphical summary tools that we have found to be necessary to simplify the otherwise arcane analysis of mixture-process experiments. As a result, JMP-specific instructions are also provided in the protocol.

An open-source R implementation of many modeling and optimization steps in this workflow is now available in the SVEMnet package\cite{karl2026svemnet}, allowing these steps to be performed without JMP Pro. The package provides SVEM fits with elastic-net and relaxed elastic-net base learners, deterministic model-term expansion, a permutation-based whole-model test heuristic, and mixture-constrained multi-response candidate scoring and selection. The JMP protocol below uses a forward-selection base learner; SVEMnet adds an independently implemented forward-selection SVEM option from version~4.0.0. The accompanying SVEMnet paper illustrates these steps with its own simulated LNP formulation example.

SVEM employs the same linear regression model foundation as the traditional approach, but it allows us to avoid tedious modifications that are required to fit a ``full model'' of candidate effects by using either a forward selection or a penalized selection (Lasso) base approach. Additionally, SVEM provides an improved ``reduced model'' fit that minimizes the potential for incorporating noise (process plus analytical variance) that appears in the data. It works by averaging the predicted models resulting from repeatedly reweighting the relative importance of each run in the model\cite{ref13,ref14,ref15,ref16,ref17,ref18}. SVEM provides a framework for modeling mixture-process experiments that is both easier to implement than traditional single-shot regression and yields better quality optimal formulation candidates\cite{ref12,ref13}. The mathematical details of SVEM are beyond the scope of this paper and even a cursory summary beyond the relevant literature review would distract from its main advantage in this application: it allows a simple, robust, and accurate click-to-run procedure for practitioners.

The presented workflow is consistent with the Quality by Design (QbD)\cite{ref19} approach to pharmaceutical development\cite{ref20}. A result of the study will be an understanding of the functional relationship that links the material attributes and process parameters to critical quality attributes (CQAs)\cite{ref21}. Daniel et al.\cite{ref22} discuss using a QbD framework specifically for RNA platform production: our workflow could be used as a tool within this framework.

\FloatBarrier
\section*{Protocol}

\subsection*{1. Recording the study purpose, responses, and factors}
\label{protocol:1}

NOTE: Throughout this protocol, JMP is used for designing and analyzing the experiment. Equivalent software can be used following similar steps. For examples and further instructions for all the steps performed in Section 1, please refer to \textbf{Supplementary File 1}.

\protocolstep{1.1}{Summarize the purpose of the experiment in a date-stamped document.}

\protocolstep{1.2}{List the primary responses (CQAs) that will be measured during the experiment.}

\protocolstep{1.3}{List any secondary responses (\textit{e.g.}, downstream restrictions on physicochemical properties) that might be measured.}

\protocolstep{1.4}{List process parameters that may be related to the responses, including those that are most relevant to the purpose of the study.}

\protocolstep{1.5}{If the study will run over multiple days, include a Day categorical ``blocking'' factor.}

NOTE: This balances factor settings across days to prevent day-level shifts in the process mean from being confounded with the study factors.

\protocolstep{1.6}{Select the factors to be varied and those to be held constant during the study.}

\begin{figure}[htbp]
\centering
\includegraphics[width=\linewidth,height=0.67\textheight,keepaspectratio]{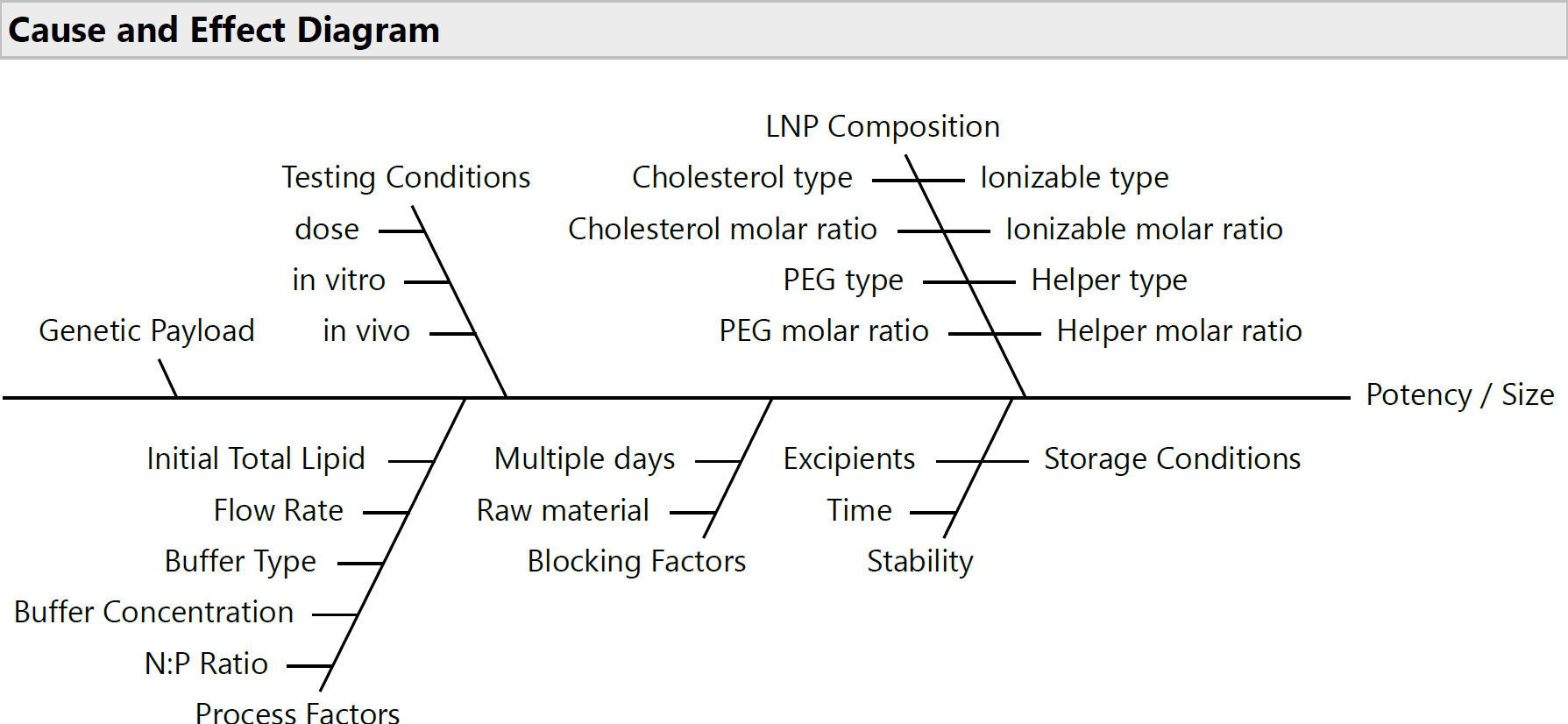}
\caption{Cause and Effect Diagram. The diagram shows common factors in an LNP formulation optimization problem.}
\label{fig:2}
\end{figure}

NOTE: Use risk prioritization tools like failure mode and effects analysis\cite{ref20} for selecting the most relevant subset of factors (\textbf{Figure~\ref{fig:2}}). Usually, all lipids should be allowed to vary, although in some budget-constrained cases, it is reasonable to lock PEG at a fixed ratio.

\protocolstep{1.7}{Establish the ranges for the varying factors and the relevant decimal precision for each.}

\protocolstep{1.8}{Decide the study design size (the number of unique batches of particles) using the minimum and maximum heuristics. Manually included control benchmark runs do not count toward the run size recommended by the heuristics.}

NOTE: The following heuristics assume the responses are continuous. The minimum heuristic assumes that it will be possible to perform a follow-up study, if needed, in addition to performing confirmation runs for candidate optimal formulations. If it will only be feasible to perform confirmation runs, then it is better to budget for the number of runs obtained from the maximum heuristic. For binary primary responses, seek help from a statistician to determine the appropriate number of runs.

\protocolstep{1.8.1}{\textbf{Minimum heuristic:} allocate three runs per mixture factor, two per continuous process factor, and one per level of each categorical factor.}

NOTE\textbf{:} For a study with four lipid factors, two continuous, and one three-level categorical process variable, this leads to a suggestion of \((3 \times 4) + (2 \times 2) + 3 = 19\) space-filling runs. Add additional runs if some are likely to fail due to formulation or measurement issues.

\protocolstep{1.8.2}{\textbf{Maximum heuristic:} Launch the software for building optimal designs and input the required parameters for a second-order model (including main effects, two-way interactions between all effects, and quadratic effects for continuous process factors). Calculate the minimum run size as per the software's algorithm. Add 1 to the result obtained from the software to define the maximum heuristic.}

NOTE: Refer to \textbf{Supplementary File 1} for detailed instructions on performing these steps. A sample case with four lipid factors, two continuous and one three-level categorical process variable, leads to a recommended run size of 34 (33 from software recommendation + 1). Any runs beyond this would likely be better used for confirmation or follow-up studies.

\subsection*{2. Creation of the design table with a space-filling design}
\label{protocol:2}

\protocolstep{2.1}{Open JMP and navigate the menu bar to \textbf{DOE \(>\) Special Purpose \(>\) Space Filling Design}.}

\protocolstep{2.2}{Enter the study responses (see \textbf{Supplementary File 1}).}

\protocolstep{2.3}{Optional: Add columns for additional responses, indicating if each is to be maximized, minimized, or targeted by clicking \textbf{Add Response}.}

NOTE: These settings can be modified later and do not affect the design. Likewise, additional columns for additional responses may be added after creating the design table.

\protocolstep{2.4}{Enter the study factors and the corresponding ranges. Use the \textbf{Mixture} button to add mixture factors, the \textbf{Continuous} button to add continuous factors, or the \textbf{Categorical} button to add categorical factors.}

\begin{figure}[htbp]
\centering
\includegraphics[width=\linewidth,height=0.67\textheight,keepaspectratio]{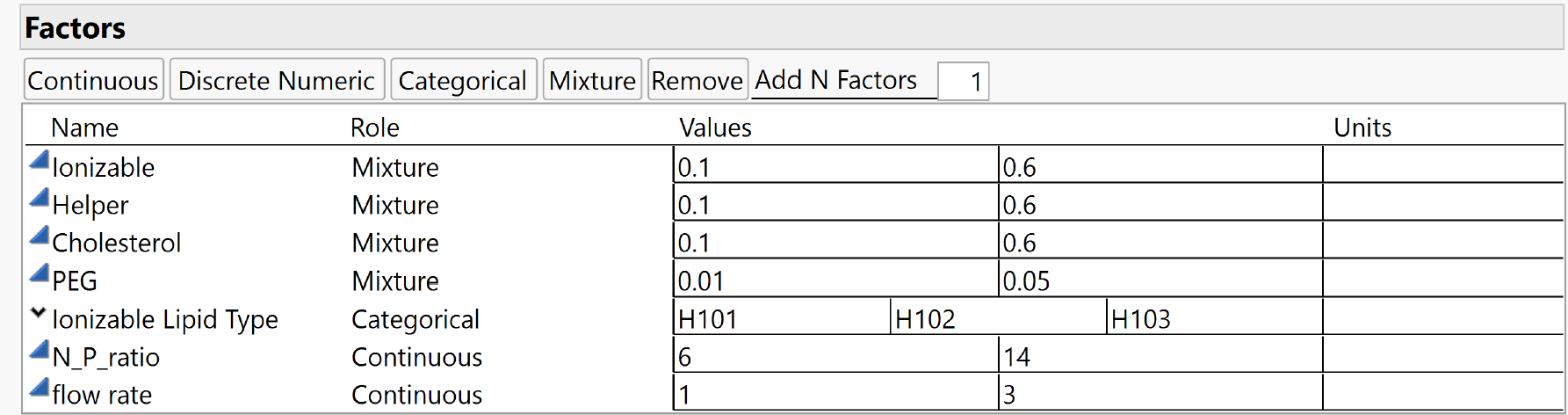}
\caption{Study factors and ranges. Screenshots of settings within experimental software are useful for reproducing the study setup.}
\label{fig:3}
\end{figure}

NOTE: This example study uses the factors and ranges illustrated in \textbf{Figure~\ref{fig:3},} which include the Ionizable molar ratio (ranging between 0.1 and 0.6), the Helper molar ratio (also between 0.1 and 0.6), the Cholesterol molar ratio (between 0.1 and 0.6), the PEG molar ratio (from 0.01 to 0.05), and the Ionizable Lipid Type (which can be H101, H102, or H103).

\protocolstep{2.5}{Input the predetermined number of runs for the design into the \textbf{Number of Runs} field.}

\protocolstep{2.6}{Optional: Increase the \textbf{Average Cluster Size} from the default of 50 to 2000 via the red triangle menu next to the \textbf{Space Filling Design} header and in the \textbf{Advanced Options} submenu.}

NOTE: This is a setting for the space-filling algorithm that can lead to slightly better design construction at the cost of additional computational time.

\protocolstep{2.7}{Generate the space-filling design table for the chosen factors and run size. Click \textbf{Fast Flexible Filling,} then click \textbf{Make Table}.}

\begin{figure}[htbp]
\centering
\includegraphics[width=\linewidth,height=0.67\textheight,keepaspectratio]{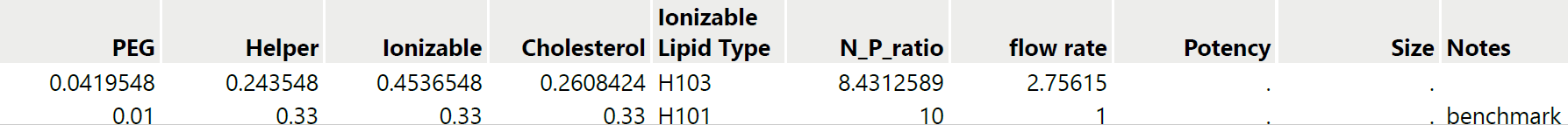}
\caption{Initial output for a space-filling design. The first two rows of the table are shown. Settings need to be rounded to the desired precision while also making sure that the lipid amounts sum to 1. The benchmark was added to the table manually.}
\label{fig:4}
\end{figure}

NOTE: The first two runs from an example design are shown in \textbf{Figure~\ref{fig:4}}.

\protocolstep{2.8}{Add a \textit{Notes} column to the table for annotating any manually created runs. Double-click the first empty column header to add a column, and then double-click the new column header to edit the name.}

\protocolstep{2.9}{If applicable, manually incorporate benchmark control runs into the design table. Include a replicate for one of the control benchmarks. Mark the benchmark name in the \textit{Notes} column and color-code the benchmark replicate rows for easy graph identification.}

\protocolstep{2.9.1}{Add a new row by double-clicking the first empty row header and input the benchmark factor settings. Duplicate this row to create a replicate of the benchmark. Highlight both rows and navigate to \textbf{Rows \(>\) Colors} to assign a color for graphing purposes.}

NOTE: The replicate provides a model-independent estimate of the process plus analytical variance and will provide additional graphical insight.

\protocolstep{2.10}{If any benchmark control runs exceed the range of the study factors, denote this in the \textit{Notes} column for future exclusion from analysis.}

\protocolstep{2.11}{Round the mixture factors to the appropriate granularity. To do so:}

\protocolstep{2.11.1}{Highlight the column headers for the mixture factors, right-click one of the column headers, and navigate to \textbf{New Formula Column \(>\) Transform \(>\) Round…}, input the correct rounding interval, and click \textbf{OK}.}

\protocolstep{2.11.2}{Make sure no rows are selected by clicking the bottom triangle at the intersection of row and column headers.}

\protocolstep{2.11.3}{Copy the values from the newly created rounded columns (\textbf{Ctrl + C}) and paste (\textbf{Ctrl + V}) into the original mixture columns. Finally, delete the temporary rounded value columns.}

\protocolstep{2.12}{After rounding the lipid ratios, verify their sum equals 100\% by selecting the column headers for the mixture factors, right-clicking one, and going to \textbf{New Formula Column \(>\) Combine \(>\) Sum}. If any row's sum does not equal 1, manually adjust one of the mixture factors, ensuring the factor setting stays within the factor range. Delete the sum column after adjustments are done.}

\protocolstep{2.13}{Follow the same procedure used for rounding the mixture factors to round the process factors to their respective granularity.}

\protocolstep{2.14}{Format the lipid columns to display as percentages with the desired number of decimals: select the column headers, right-click, and choose \textbf{Standardize Attributes...}. In the next window, set \textbf{Format} to \textbf{Percent} and adjust the number of decimals as needed.}

\protocolstep{2.15}{If manual runs are added such as benchmarks, re-randomize the table row order: add a new column with random values (Right-click the last column header and select \textbf{New Formula Column \(>\) Random \(>\) Random Normal}). Sort this column in ascending order by right-clicking on its column header, and then delete the column.}

\protocolstep{2.16}{Optional: add a \textit{Run ID} column. Populate this with the current date, experiment name, and row number from the table.}

\begin{figure}[htbp]
\centering
\includegraphics[width=\linewidth,height=0.67\textheight,keepaspectratio]{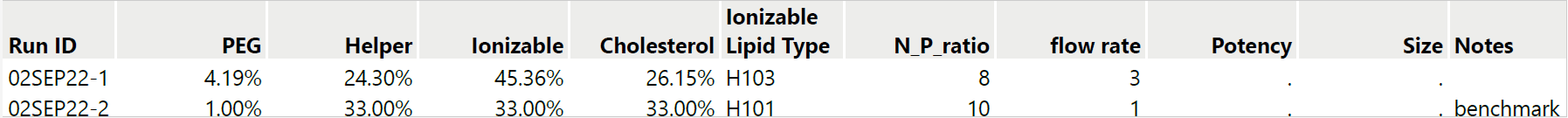}
\caption{Formatted study table. The factor levels have been rounded and formatted and a Run ID column has been added.}
\label{fig:5}
\end{figure}

NOTE: See \textbf{Figure~\ref{fig:5}} for an example.

\begin{figure}[htbp]
\centering
\includegraphics[width=\linewidth,height=0.67\textheight,keepaspectratio]{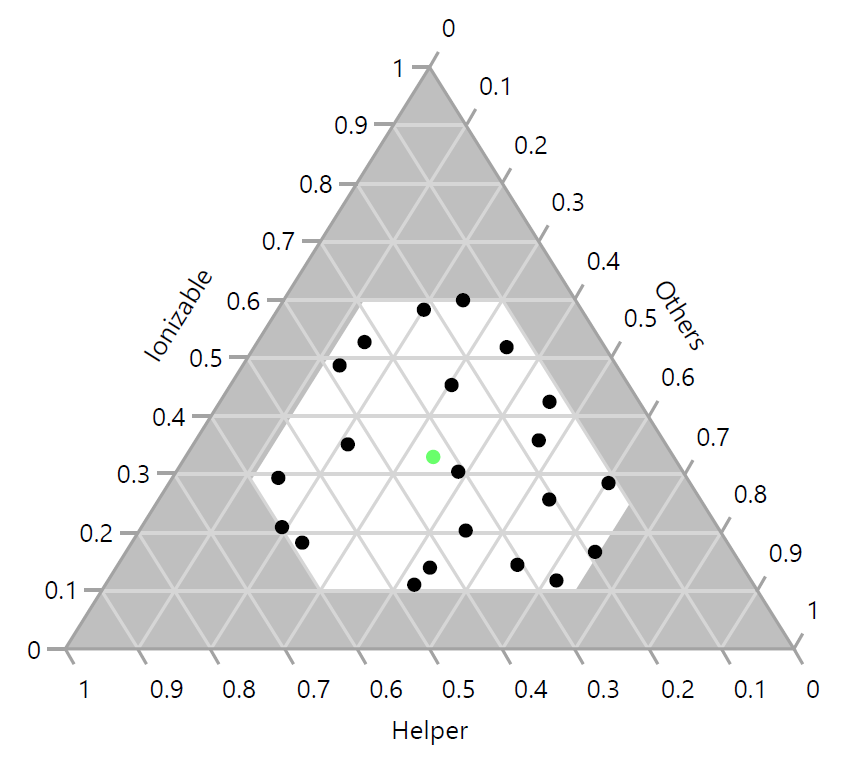}
\caption{Design points on a ternary plot. The 23 formulations are shown as a function of the corresponding Ionizable, Helper and ``Others'' (Cholesterol+PEG) ratios. The green point in the center represents the benchmark 33:33:33:1 molar ratio of Ionizable (H101):Cholesterol:Helper (DOPE):PEG.}
\label{fig:6}
\end{figure}

\begin{figure}[htbp]
\centering
\includegraphics[width=\linewidth,height=0.67\textheight,keepaspectratio]{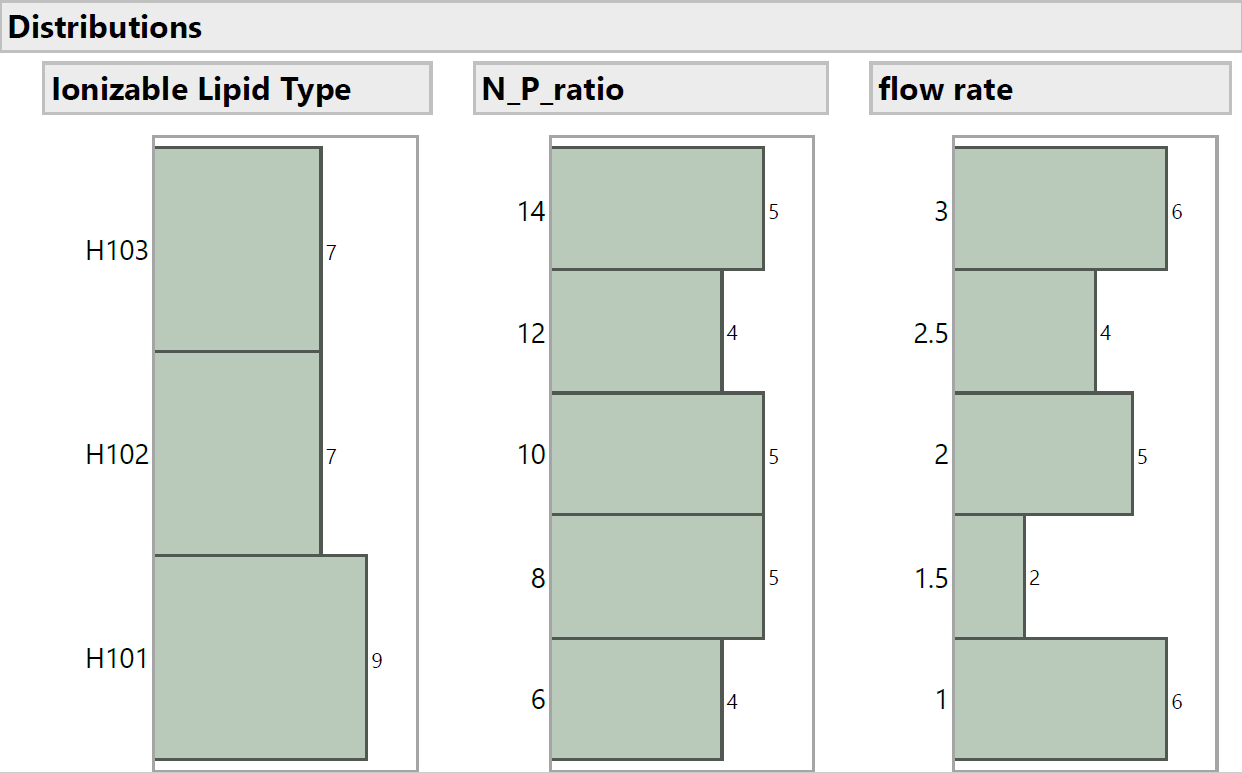}
\caption{Distribution of non-mixture process factors in the experiment. The histograms show how the experimental runs are spaced across Ionizable Lipid Type, N:P ratio, and Flow Rate.}
\label{fig:7}
\end{figure}

\protocolstep{2.17}{Generate ternary plots to visualize the design points over the lipid factors (\textbf{Figure~\ref{fig:6}}): select \textbf{Graph \(>\) Ternary Plot} and select only the mixture factors for \textbf{X, Plotting}. Also examine the run distribution over the process factors (\textbf{Figure~\ref{fig:7}}).}

\protocolstep{2.17.1}{To examine the distribution over the process factors, select \textbf{Analyze \(>\) Distribution} and enter the process factors for \textbf{Y, Columns}.}

NOTE: The formulation scientist should confirm the feasibility of all runs. If infeasible runs exist, restart the design considering the newly discovered constraints.

\subsection*{3. Running the experiment}
\label{protocol:3}

\protocolstep{3.1}{Run the experiment in the order provided by the design table. Record the readouts in the columns built into the experimental table.}

\protocolstep{3.2}{If multiple assays are performed for the same response on an identical formulation batch, calculate an average for these results within each batch. Add a column for each assay measurement to the table.}

\protocolstep{3.2.1}{To obtain an average, select all related columns, right-click on one of the selected column headers, and choose \textbf{New Formula Column \(>\) Combine \(>\) Average}. Use this \textit{Average} column for future response analysis.}

NOTE: Without starting the recipe anew, repeated assay measurements only capture assay variance and do not constitute independent replicates.

\protocolstep{3.3}{Document any occurrence of formulation precipitation or \textit{in vivo} tolerability issues (such as severe body weight loss or death) with binary (0/1) indicators in a new column for each type of issue.}

\subsection*{4. Analyzing the experimental results}
\label{protocol:4}

\protocolstep{4.1}{Plot the readings and examine the distributions of the responses: open \textbf{Graph \(>\) Graph Builder} and drag each response into the \textbf{Y} area for individual plots. Repeat this for all responses.}

\begin{figure}[htbp]
\centering
\includegraphics[width=0.65\linewidth,height=0.43\textheight,keepaspectratio]{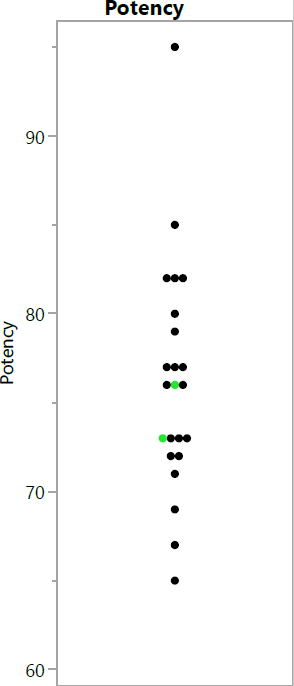}
\caption{Observed potency readings from the experiment. The points show the potency values that were observed from the 23 runs; the replicated benchmark runs are shown in green.}
\label{fig:8}
\end{figure}

\protocolstep{4.2}{Examine the relative distance between the color-coded replicate runs, if one was included. This allows for the understanding of the total (process and analytic) variation at the benchmark compared to the variability due to changes in the factor settings across the entire factor space (\textbf{Figure~\ref{fig:8}}).}

\protocolstep{4.3}{Determine if the raw response should be modeled or if a transformation should be used instead. For responses that are restricted to being positive but are unbounded above (\textit{e.g.}, potency), fit both a normal distribution and a lognormal distribution to the experimental results. If the lognormal distribution fits better with a lower AICc (corrected Akaike's Information Criterion), consider modeling that response on the log scale.}

\protocolstep{4.3.1}{Navigate to \textbf{Analyze \(>\) Distribution} and select the response for \textbf{Y, Columns}. In the resulting distribution report, click the red triangle next to the response name and choose \textbf{Continuous Fit \(>\) Fit Normal} and \textbf{Continuous Fit \(>\) Fit Lognormal} from the drop-down menu. In the subsequent \textbf{Compare Distributions} report, check the AICc values to ascertain which distribution fits the response better.}

\protocolstep{4.3.2}{To perform a log transform, right-click the response column header and select \textbf{New Formula Column \(>\) Log \(>\) Log}. When a model is built and a prediction column on the log scale is saved, transform the saved predictions back to the original scale by selecting \textbf{New Formula Column \(>\) Log \(>\) Exp}. }

\protocolstep{4.3.3}{For proportion responses strictly between 0 and 1, compare the fit of a normal and beta distribution. If the beta distribution has a lower AICc, consider a logit transform. In the Distribution report for the response, choose \textbf{Continuous Fit \(>\) Fit Normal} and \textbf{Continuous Fit \(>\) Fit Beta}.}

\protocolstep{4.3.3.1}{For the logit transform, right-click the response column header in the data table, and select \textbf{New Formula Column \(>\) Specialty \(>\) Logit}. Post model-building, save the prediction column. To revert to the original scale, use \textbf{New Formula Column \(>\) Specialty \(>\) Logistic}.}

NOTE: The pooled (marginal) distributions offer practitioners a quick first look at the response scale before fitting a model. However, differences among recipe means can skew the pooled readings even when scatter at a given recipe is roughly constant. Use scientific and assay knowledge, replicate preparations and, when informative, fitted residuals to guide the choice. For positive responses, roughly constant absolute scatter favors original units, whereas proportional scatter can favor logs and fold-change interpretation; a log scale may suit firefly luciferase (FLuc) readings when assay knowledge supports proportional scatter. Clarify the meaning of zeros before considering logs. A lower pooled AICc alone does not establish that logs or logits are appropriate.

\protocolstep{4.4}{Graph the runs on a ternary plot. Color the points according to the responses (or the transformed responses if a transformation was applied): open \textbf{Graph \(>\) Ternary Plot}. Select only the mixture factors for \textbf{X, Plotting}. Right-click on any of the resulting graphs, select \textbf{Row Legend} and then select the (transformed) response column.}

NOTE: Coloring the points according to responses gives a model-independent visual perspective of behavior in relation to mixture factors.

\protocolstep{4.5}{Delete the \textit{Model} script generated by the Space-Filling Design.}

\protocolstep{4.6}{Build an independent model for each response as a function of the study factors, repeating the following steps for each response.}

NOTE: In the case of a secondary binary response (e.g., formulation failure or mouse death), model this response as well. Change the target distribution setting from \textbf{Normal} to \textbf{Binomial}.

\protocolstep{4.7}{Construct a ``full'' model comprising all candidate effects. This model should include the main effects of each factor, two- and three-way interactions, quadratic and partial cubic terms in the process factors, and Scheffé cubic terms for the mixture factors\cite{ref23,ref24}.}

\begin{figure}[htbp]
\centering
\includegraphics[width=\linewidth,height=0.67\textheight,keepaspectratio]{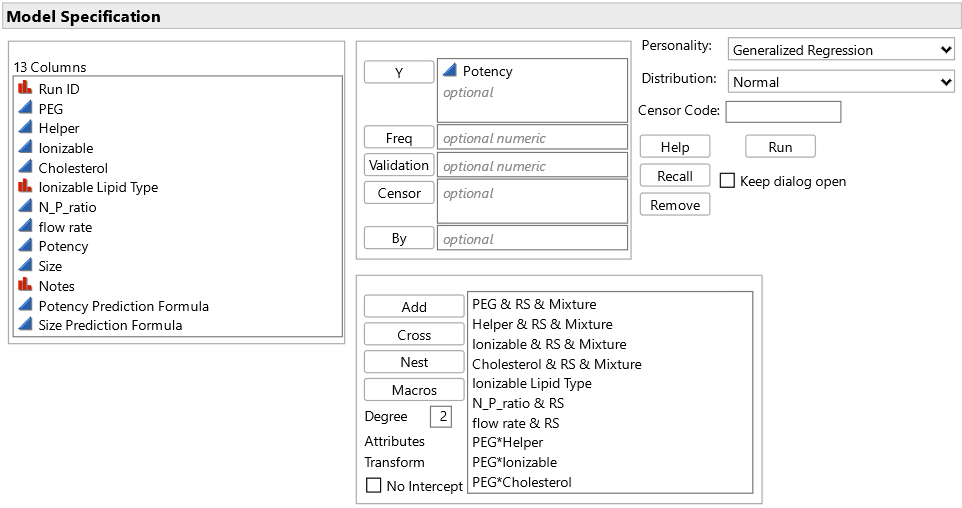}
\caption{Software dialog for initiating the analysis. The candidate effects have been entered along with the target responses, and the No Intercept option has been unchecked.}
\label{fig:9}
\end{figure}

NOTE: Use the same set of candidate effects for each response. The SVEM model selection technique will independently refine the models for each response, potentially resulting in unique reduced models for each one. \textbf{Figure~\ref{fig:9}} illustrates some of these candidate effects. The following sub-steps detail this process.

\protocolstep{4.7.1}{Select \textbf{Analyze \(>\) Fit Model}.}

\protocolstep{4.7.2}{Ensure blocking factors (\textit{e.g.}, Day) are not allowed to interact with other study factors. Select any blocking factors and click \textbf{Add}. Do not include these factors in any of the subsequent sub-steps.}

NOTE: Blocking factors are important to account for in the model but blocking factors should not be allowed to interact with other study factors. The main purpose of blocking factors is to help control the experiment's variability and improve the sensitivity of the experiment.

\protocolstep{4.7.3}{Highlight all of the study factors. Modify the \textbf{Degree} field value to 3 (it is set to 2 by default). Click on \textbf{Factorial to Degree}.}

NOTE: This action includes main effects as well as two- and three-way interactions in the model.

\protocolstep{4.7.4}{Select only the non-mixture factors in the selection window. Click \textbf{Macros \(>\) Partial Cubic}.}

NOTE: This action introduces quadratic effects for the continuous process factors and the interaction of the continuous process factors with other non-mixture factors in the model.

\protocolstep{4.7.5}{Choose only the mixture factors from the selection list. Click \textbf{Macros \(>\) Scheffe Cubic}. Deactivate the default \textbf{No Intercept} option (refer to \textbf{Figure~\ref{fig:9}}).}

NOTE: Including an intercept in the model is an essential step when using Lasso methods and is also helpful in the forward selection context. The traditional default setting \textbf{No Intercept} is usually in place because fitting an intercept simultaneously with all the mixture main effects, without modifications such as the SVEM approach, is not feasible with the regular least squares regression procedure\cite{ref12}.

\protocolstep{4.7.6}{Specify the response column: highlight the response column and click \textbf{Y}.}

\protocolstep{4.7.7}{Change the \textbf{Personality} setting to \textbf{Generalized Regression}. Keep \textbf{Distribution} set to \textbf{Normal}.}

\protocolstep{4.7.8}{Save this model setup to the data table for use with additional responses by clicking on the red triangle menu next to \textbf{Model Specification} and selecting \textbf{Save to Data Table}.}

\protocolstep{4.8}{Apply the SVEM forward selection method to fit the reduced model, without mandatory inclusion of the mixture factor main effects, and store the prediction formula column in the data table.}

\protocolstep{4.8.1}{From the \textbf{Fit Model} dialog, click \textbf{Run}.}

\protocolstep{4.8.2}{For the \textbf{Estimation Method}, select \textbf{SVEM Forward Selection}.}

\begin{figure}[htbp]
\centering
\includegraphics[width=0.5\linewidth,height=0.44\textheight,keepaspectratio]{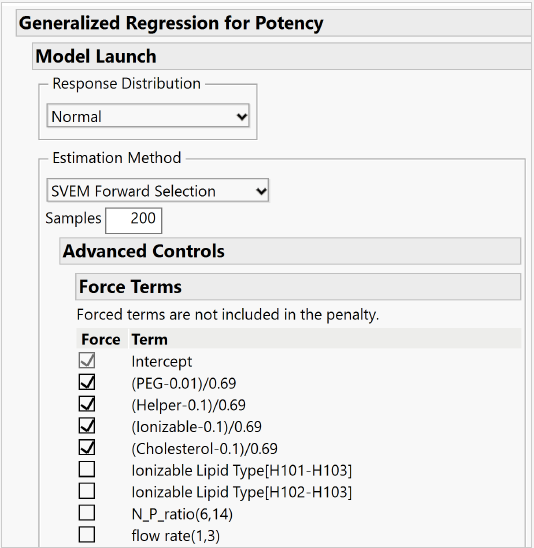}
\caption{Additional dialog for specifying SVEM options. By default, the lipid main effects are forced into the model. Because an intercept is included, we recommend unchecking these boxes in order not to force the effects.}
\label{fig:10}
\end{figure}

\protocolstep{4.8.3}{Expand the \textbf{Advanced Controls \(>\) Force Terms} menus and deselect the boxes associated with the mixture factors' main effects. Only the \textbf{Intercept} term box should remain checked. \textbf{Figure~\ref{fig:10}} displays the default setup where the main effects are forced. For this step, these boxes need to be unchecked to allow the model to include or exclude these effects based on the forward selection procedure.}

\protocolstep{4.8.4}{Click \textbf{Go} to run the SVEM Forward Selection procedure.}

\begin{figure}[htbp]
\centering
\includegraphics[width=\linewidth,height=0.67\textheight,keepaspectratio]{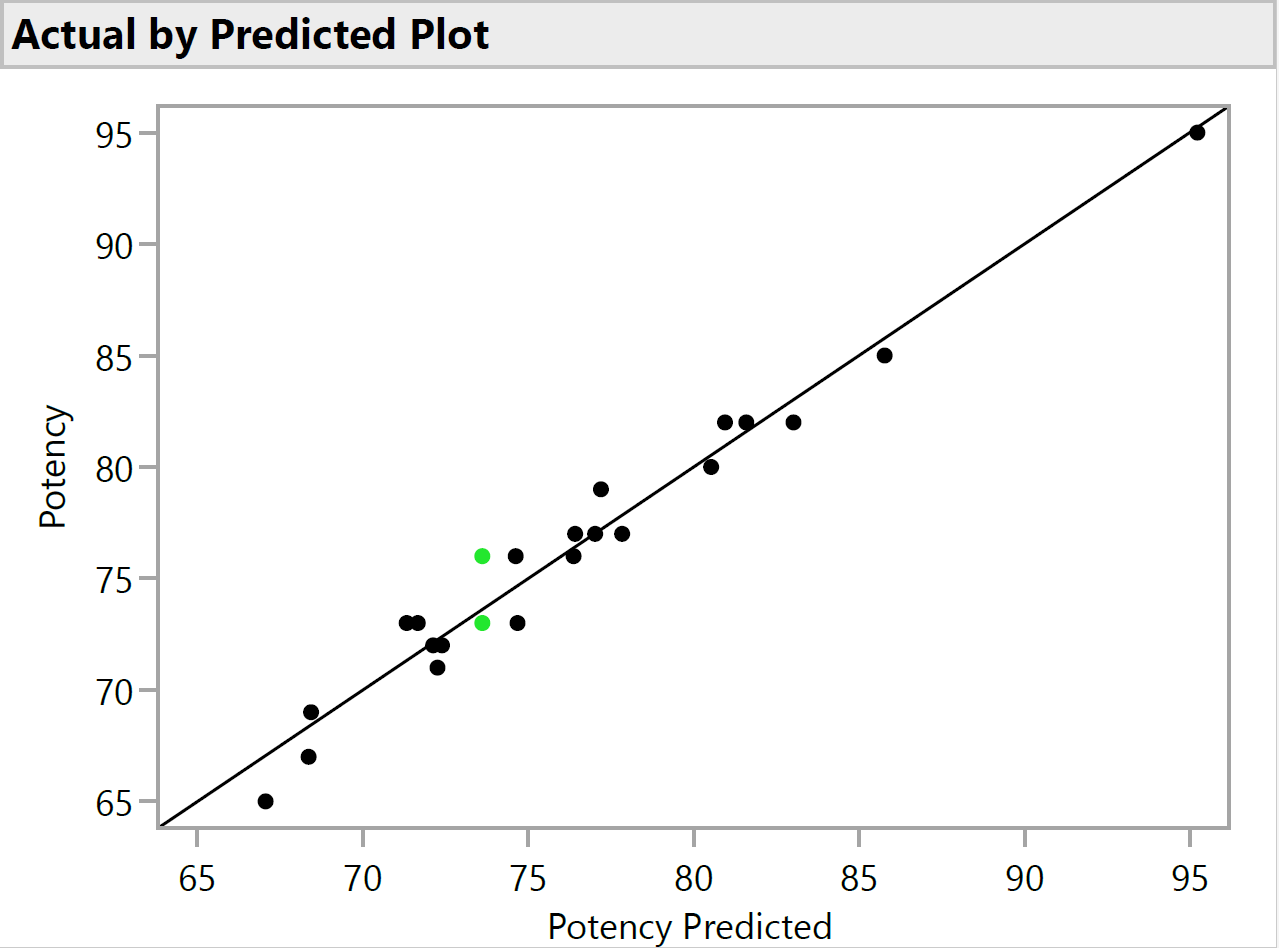}
\caption{Actual by predicted plot. This figure plots the observed Potency against the value predicted for each formulation by the SVEM model. These are fitted values for the training runs; inspect the relationship and check for outliers, while recognizing that correlation alone does not establish predictive signal.}
\label{fig:11}
\end{figure}

\protocolstep{4.9}{Plot the actual responses against the predicted responses from the SVEM model to inspect the fitted relationship and check for outliers (\textbf{Figure~\ref{fig:11}}).}

NOTE: A close fit in a training actual-by-predicted plot does not by itself establish predictive signal. In later work, Karl\cite{karl2024wholemodel} proposed a permutation-based whole-model test heuristic for SVEM and illustrated it with a simulated scenario based on this LNP example, with an additional PDI response. Because PDI was generated independently of the study factors, its near-perfect actual-by-predicted fit was spurious. Consistent with this, the test did not reject a constant response surface for PDI (approximate whole-model \(p\)-values of 0.9180 with SVEM Forward Selection and 0.9733 with SVEM Lasso), whereas Potency and Size showed strong evidence of nonconstant response surfaces. These diagnostics complement subject-matter judgment and confirmation runs; a nonsignificant test alone does not establish the absence of a relationship.

\protocolstep{4.9.1}{Click the red triangle next to \textbf{SVEM Forward Selection} and select \textbf{Diagnostic Plots \(>\) Plot Actual by Predicted}.}

\protocolstep{4.9.2}{Click the red triangle next to \textbf{SVEM Forward Selection} and select \textbf{Save Columns \(>\) Save Prediction Formula} to create a new column containing the prediction formula in the data table.}

\protocolstep{4.9.3}{Optional: repeat the above steps using \textbf{SVEM Lasso} as the \textbf{Estimation Method} to determine if a different optimal recipe is suggested after performing the subsequent steps. If so, run both recipes as confirmation runs (discussed in Section 5) to see which performs best in practice\cite{ref12}.}

\protocolstep{4.10}{Repeat the model-building steps for each response.}

\begin{figure}[htbp]
\centering
\includegraphics[width=\linewidth,height=0.67\textheight,keepaspectratio]{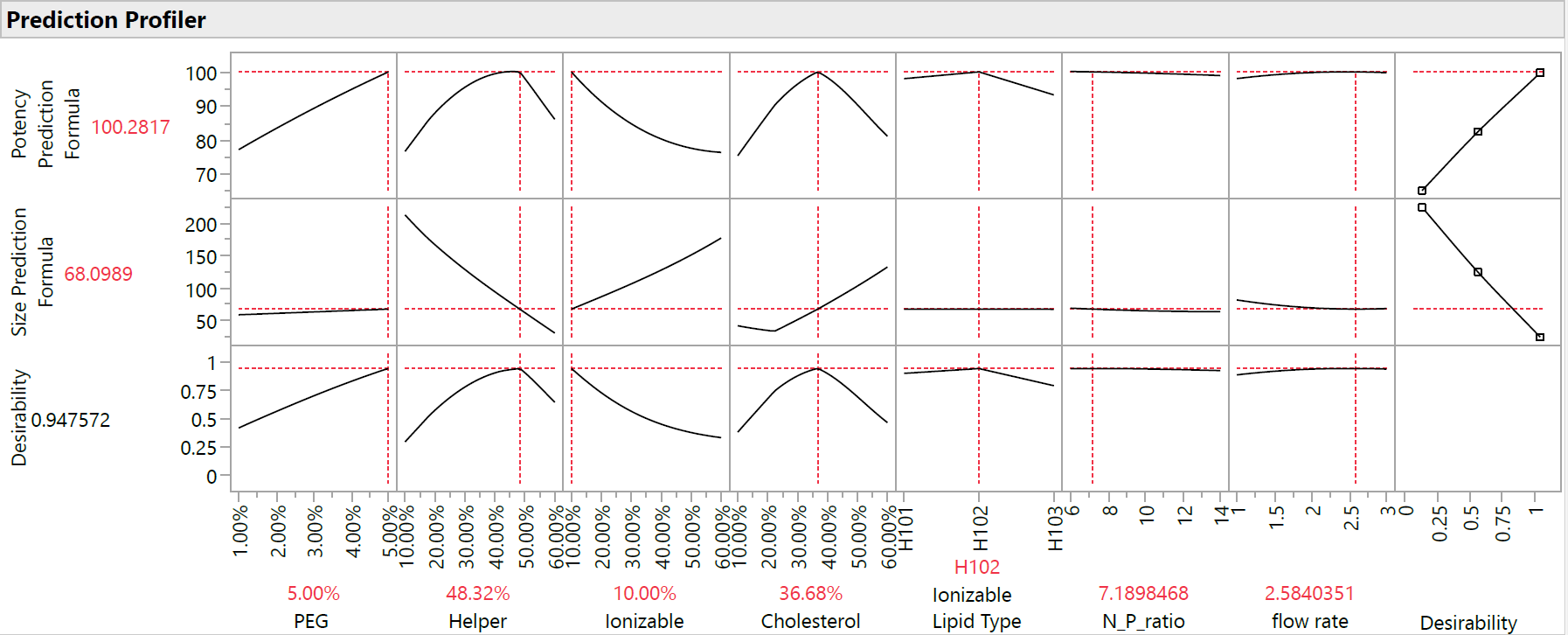}
\caption{Prediction profiler. The top two rows of graphs show the slices of the predicted response function at the optimum formulation (as identified by the SVEM approach). The bottom row of graphs shows the weighted ``desirability'' of the formulation, which is computed from the desirability functions in the last column of graphs; these show that Potency should be maximized and Size minimized.}
\label{fig:12}
\end{figure}

\protocolstep{4.11}{Once prediction columns for all responses are saved to the data table, graph the response traces for all predicted response columns using the Profiler platform: Select \textbf{Graph \(>\) Profiler}, and select all of the prediction columns created in the previous step for \textbf{Y, Prediction Formula}, and click \textbf{OK} (\textbf{Figure~\ref{fig:12}}).}

\protocolstep{4.12}{Identify candidate optimal formulation(s).}

\protocolstep{4.12.1}{Define the ``desirability function'' for each response, specifying whether the response should be maximized, minimized, or matched to a target. Set any primary responses to use an importance weight of 1.0 and any secondary responses to use an importance weight of 0.2. From the \textbf{Prediction Profiler} red triangle menu, select \textbf{Optimization and Desirability \(>\) Desirability Functions,} then \textbf{Optimization and Desirability \(>\) Set Desirabilities}. Enter the settings into the subsequent windows.}

NOTE: The importance weights are relative and subjective, so it is worth checking the sensitivity of the combined optimum to changes in these weights within a reasonable range (\textit{e.g.}, from equal weighting to 1:5 weighting).

\protocolstep{4.12.2}{Command the Profiler to find the optimal factor settings that maximize the desirability function (\textbf{Figure~\ref{fig:12}}): from the Profiler, select \textbf{Optimization and Desirability \(>\) Maximize Desirability}.}

NOTE: The predicted values of the responses at the optimal candidates may overestimate the value of right-skewed responses such as potency; however, the confirmation runs will provide more accurate observations at these candidate formulations. The main aim is to \textit{locate} the optimal formulation (the \textit{settings} of the optimal recipe).

\protocolstep{4.12.3}{Record the optimal factor settings and note the importance weights used for each response: from the \textbf{Prediction Profiler} menu, select \textbf{Factor Settings \(>\) Remember Settings.}}

\protocolstep{4.13}{Optional: for categorical factors such as ionizable lipid type, find the conditionally optimal formulations for each factor level.}

\protocolstep{4.13.1}{First set the desired level of the factor in the profiler, then hold the \textit{Alt} key and left-click inside the graph of that factor and select \textbf{Lock Factor Setting}. Select \textbf{Optimization and Desirability \(>\) Maximize Desirability} to find the conditional optimum with this factor locked at its current setting.}

\protocolstep{4.13.2}{Unlock the factor settings before proceeding, using the same menu used to lock the factor settings.}

\begin{figure}[htbp]
\centering
\includegraphics[width=\linewidth,height=0.67\textheight,keepaspectratio]{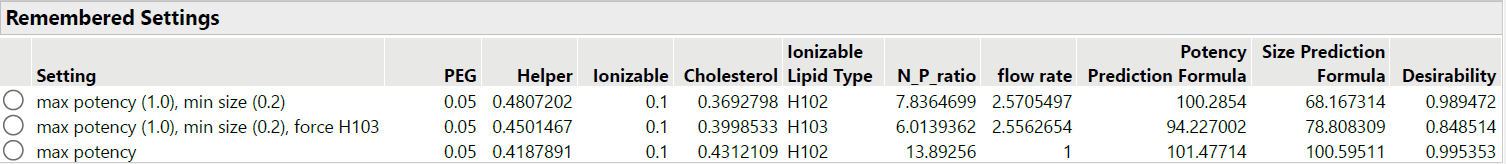}
\caption{Three optimal formulation candidates from SVEM-Forward Selection. Changing the relative importance weighting of the responses can lead to different optimal formulations.}
\label{fig:13}
\end{figure}

\protocolstep{4.14}{Repeat the optimization process after adjusting the importance weights of the responses (using \textbf{Optimization and Desirability \(>\) Set Desirabilities}), perhaps only optimizing the primary response(s) or setting some of the secondary responses to have more or less importance weight, or by setting the goal of the secondary responses to \textbf{None} (\textbf{Figure~\ref{fig:13}}).}

\protocolstep{4.15}{Record the new optimal candidate (from the \textbf{Prediction Profiler} menu, select \textbf{Factor Settings \(>\) Remember Settings}).}

\protocolstep{4.16}{Produce graphical summaries of the optimal regions of the factor space: generate a data table with 50,000 rows populated with randomly generated factor settings within the allowed factor space, along with the corresponding predicted values from the reduced model for each of the responses and the joint desirability function.}

\protocolstep{4.16.1}{In the Profiler, select \textbf{Output Random Table}. Set \textbf{How many runs to simulate?} to 50,000 and click \textbf{OK}.}

NOTE: This generates a new table with the predicted values of the responses at each of the 50,000 formulations. The \textbf{Desirability} column depends on the importance weights for the responses that are in place when the \textbf{Output Random Table} option is selected.

\protocolstep{4.16.2}{In the newly created table, add a new column that calculates the percentile of the \textit{Desirability} column. Use this percentile column in the ternary plots instead of the raw \textit{Desirability} column. Right-click the \textit{Desirability} column header and select \textbf{New Formula Column \(>\) Distributional \(>\) Cumulative Probability} to create a new \textit{Cumulative Probability[Desirability]} column.}

\protocolstep{4.16.3}{Generate the graphics described in the following steps. Repeatedly alter the color scheme of the graphics in order to display the predictions for each response and for the \textit{Cumulative Probability[Desirability]} column.}

\protocolstep{4.16.4}{Construct ternary plots for the four lipid factors. In the table, navigate to \textbf{Graph \(>\) Ternary Plot}, select the mixture factors for \textbf{X, Plotting,} and click \textbf{OK}. Right-click in one of the resulting graphs, select \textbf{Row Legend}, and then select the predicted response column. Change the \textbf{Colors} dropdown to \textbf{Jet}.}

\begin{figure}[htbp]
\centering
\includegraphics[width=\linewidth,height=0.67\textheight,keepaspectratio]{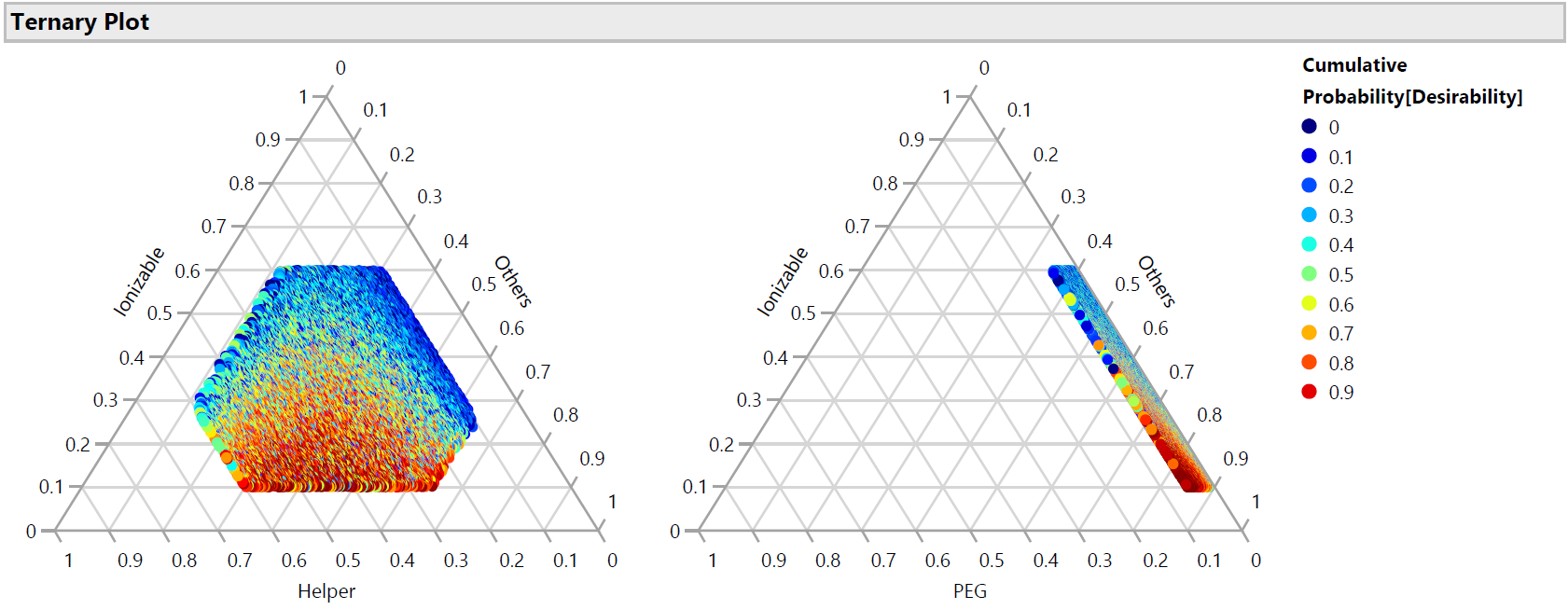}
\caption{Ternary plots for the percentile of desirability. The plot shows the 50,000 formulations color coded by percentile of desirability, where the desirability is set with an importance weight of 1.0 for maximizing Potency and 0.2 for minimizing size. These plots show that the optimal region of formulations consists of lower percentages of ionizable lipid and higher percentages of PEG.}
\label{fig:14}
\end{figure}

\begin{figure}[htbp]
\centering
\includegraphics[width=\linewidth,height=0.67\textheight,keepaspectratio]{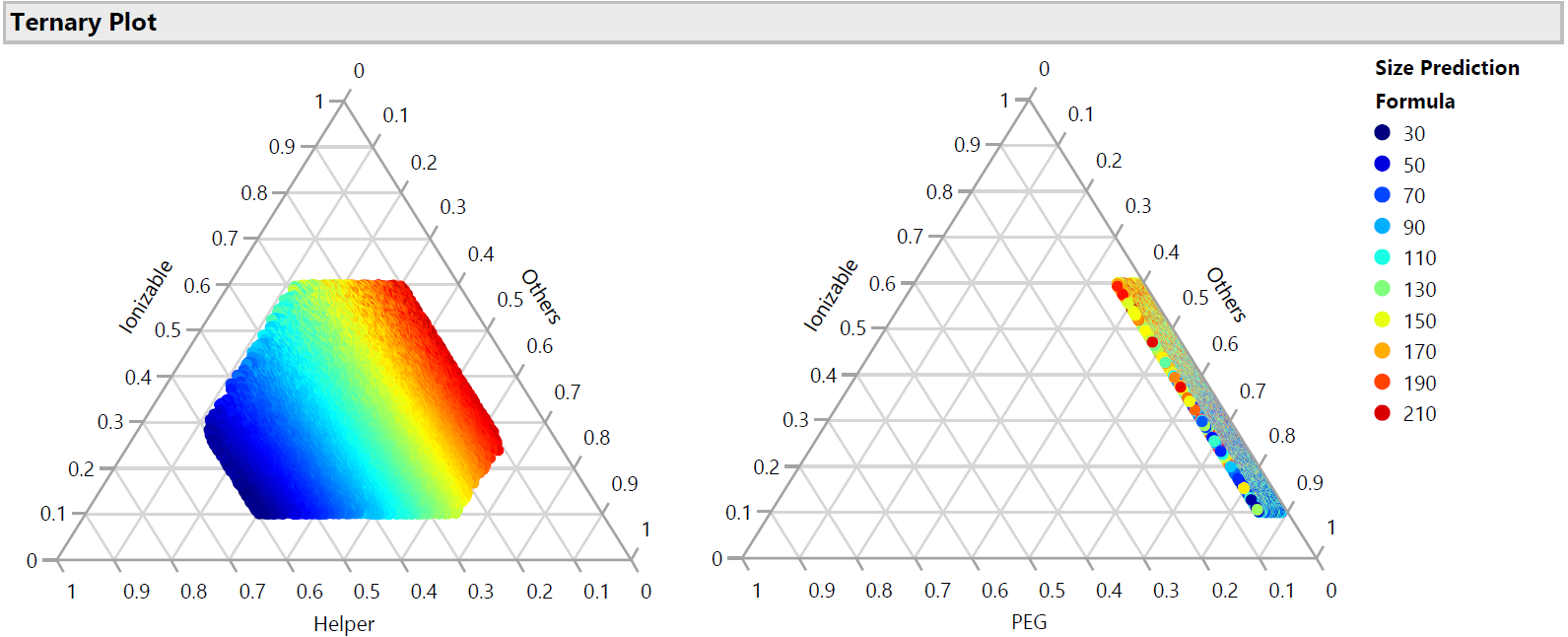}
\caption{Ternary plot for the predicted Size. The plot shows the size predictions from the SVEM model for each of the 50,000 formulations. Size is minimized with higher percentages of helper lipid and maximized with lower percentages of helper. Since the other factors vary freely across the 50,000 plotted formulations, this implies that this relationship holds across the ranges of the other factors (PEG, flow rate, etc.).}
\label{fig:15}
\end{figure}

NOTE: This displays the best and worst performing regions with respect to the lipid factors. \textbf{Figure~\ref{fig:14}} shows the percentiles of the joint desirability when considering maximizing \textit{Potency} (importance=1) and minimizing \textit{Size} (importance=0.2), while averaging over any factors that are not shown on the ternary plot axes. \textbf{Figure~\ref{fig:15}} shows the raw predicted size. It is also reasonable to break down these graphs conditionally on other factors, such as creating a distinct set of ternary plots for each ionizable lipid type with a \textbf{Local Data Filter} (available from the red triangle menu next to \textbf{Ternary Plot}).

\protocolstep{4.16.5}{Similarly, use \textbf{Graph \(>\) Graph Builder} to plot the 50,000 color-coded points (representing unique formulations) against the non-mixture process factors, either individually or jointly, and search for relationships between the response(s) and the factor(s). Look for the factor settings that yield the highest desirability. Explore different combinations of factors in the graphics.}

\begin{figure}[htbp]
\centering
\includegraphics[width=\linewidth,height=0.67\textheight,keepaspectratio]{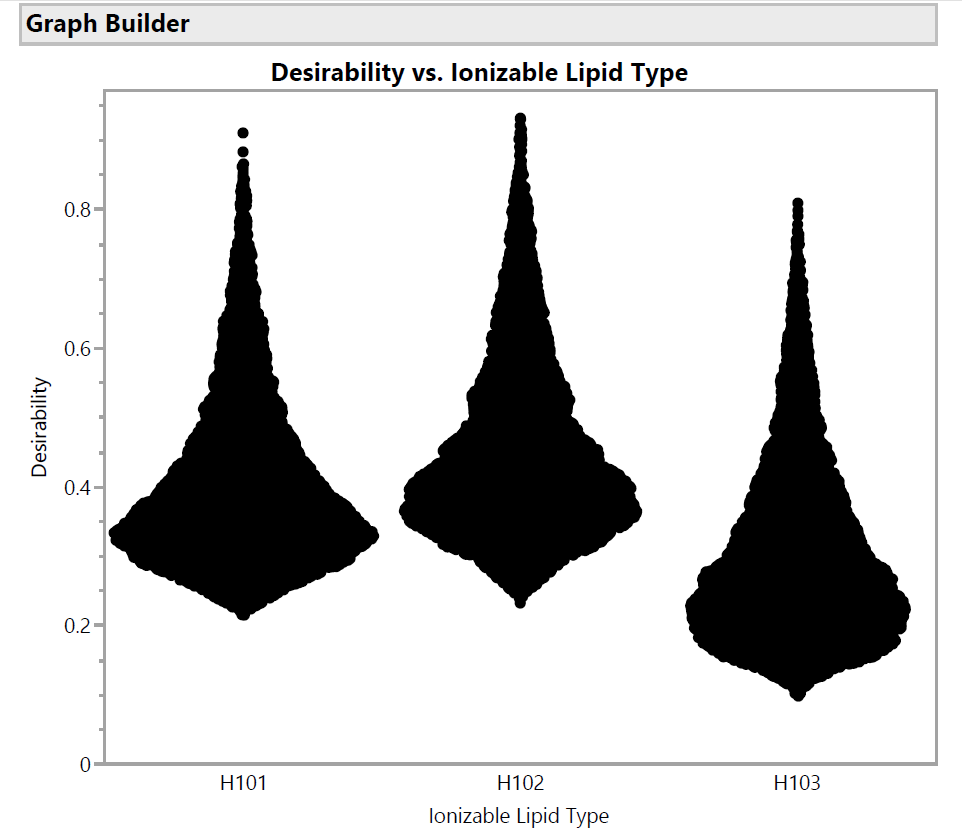}
\caption{Violin plots for the desirability of formulations involving the three different ionizable lipid types. Each of the 50,000 points represents a unique formulation from throughout the allowed factor space. The upper endpoints show the highest predicted desirability attained among the sampled recipes for each type. H102 attains the highest sampled desirability. SVEM selects or shrinks candidate effects for prediction, and this graph compares the predicted practical differences across the factor levels.}
\label{fig:16}
\end{figure}

NOTE: When coloring graphs, use \textit{Cumulative Probability[Desirability],} but when plotting the desirability on the vertical axis against process factors use the raw \textit{Desirability} column. The \textit{Desirability} column can also be placed on an axis of the \textbf{Graph \(>\) Scatterplot 3D} visualization along with two other process factors for multivariate exploration. \textbf{Figure~\ref{fig:16}} shows the joint desirability of the sampled formulations for each of the three ionizable lipid types. The most desirable formulations use H102, with H101 providing some potentially competitive alternatives.

\protocolstep{4.16.6}{Save the Profiler and its remembered settings back to the data table. Click the red triangle next to \textbf{Profiler} and select \textbf{Save Script \(>\) To Data Table…}.}

\subsection*{5. Confirmation Runs}
\label{protocol:5}

\begin{figure}[htbp]
\centering
\includegraphics[width=\linewidth,height=0.67\textheight,keepaspectratio]{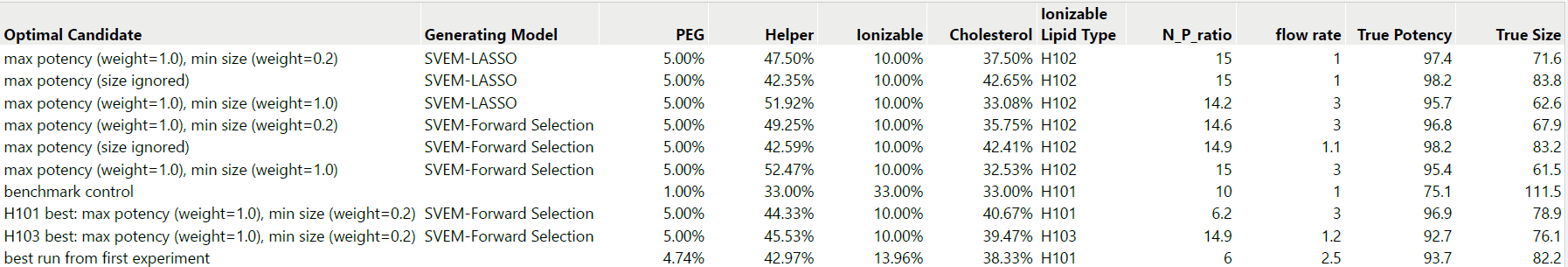}
\caption{Table of ten optimal candidates to be run as confirmation runs. The True Potency and True Size have been filled in from the simulation generating functions (without any added process or analytical variation).}
\label{fig:17}
\end{figure}

\protocolstep{5.1}{Prepare a table listing the optimal candidates identified previously (\textbf{Figure~\ref{fig:17}}).}

NOTE: The \textit{True Potency} and \textit{True Size} values in \textbf{Figure~\ref{fig:17}} are filled in using the simulated generating functions: in actual practice, these will be approximated by formulating and then measuring the performance of these recipes.

\protocolstep{5.1.1}{Include the benchmark control with the set of candidate runs that will be formulated and measured.}

\protocolstep{5.1.2}{If any of the formulations from the experiment were found to yield desirable results, perhaps by outperforming the benchmark, select the best to add to the candidate table and retest along with new formulations.}

NOTE: Either manually add desired runs to the candidate table or use the Profiler window's \textbf{Remembered Settings} if these runs are from the previous experiment. Identify the run's row number, navigate to \textbf{Prediction Profiler \(>\) Factor Settings \(>\) Set to Data in Row}, and input the row number. Then, choose \textbf{Prediction Profiler \(>\) Factor Settings \(>\) Remember Settings} and label appropriately (\textit{e.g.}, ``benchmark'' or ``best run from previous experiment'').

\protocolstep{5.1.3}{Right-click on the \textbf{Remembered Settings} table in the Profiler and select \textbf{Make into Data Table}.}

NOTE: Depending on the study's priority and budget, consider running replicates for each confirmation run, especially if replacing the benchmark. Create and analyze each formulation twice, using the average result for ranking. Pay attention to any candidates with a wide response range across the two replicates as this could indicate high process variance.

\protocolstep{5.1.4}{If necessary due to budget constraints, downselect from the identified candidates to match the experimental budget or to eliminate redundant candidates.}

\protocolstep{5.2}{Carry out the confirmation runs. Construct the formulations and gather the readouts.}

\protocolstep{5.3}{Check for consistency between the results from the original experiment and the results for the confirmation batch for benchmarks or other repeated recipes. If there is a large and unexpected shift, then consider what might have contributed to the shift and if it is possible that all runs from the confirmation batch were affected.}

\protocolstep{5.4}{Compare the performance of the candidate optimal formulations. Explore whether any new candidates outperformed the benchmark.}

\protocolstep{5.5}{Optional: add the result of the confirmation runs to the experimental table and rerun the analysis in Section 4.}

NOTE: The next step of the workflow provides instructions for constructing a follow-up study along with these runs, if desired.

\subsection*{6. Optional: Designing a follow-up study to be run concurrently with the confirmation runs}
\label{protocol:6}

\protocolstep{6.1}{Assess the need for a follow-up study considering the following criteria:}

\protocolstep{6.1.1}{Determine if the optimal formulation lies along one of the factor boundaries and if a second experiment is desired to expand at least one of the factor ranges.}

\protocolstep{6.1.2}{Evaluate if the initial experiment used a relatively small run size or relatively large factor ranges and if there is a need to ``zoom in'' on the identified optimal region with additional runs and updated analysis.}

\protocolstep{6.1.3}{Check if an additional factor is being introduced. This could be a level of a categorical factor such as an additional ionizable lipid or a factor that remained constant in the initial study, for example, buffer concentration.}

\protocolstep{6.1.4}{If none of the above conditions are met, proceed to Section 7.}

\protocolstep{6.2}{Prepare for additional experimental runs to be carried out concurrently with the confirmation runs.}

\protocolstep{6.2.1}{Define the factor limits ensuring a partial overlap with the region from the initial study. If no overlap exists, a new study must be designed.}

\protocolstep{6.2.2}{Develop the new experimental runs with a space-filling design. Select \textbf{DOE \(>\) Special Purpose \(>\) Space Filling Design}.}

NOTE: For advanced users, consider a D-optimal design via \textbf{DOE \(>\) Custom Design}.

\protocolstep{6.2.3}{After the space-filling runs are generated, manually incorporate two or three runs from the original experiment that lie within the new factor space. Distribute these runs randomly within the experimental table using the steps described in Section 2 to add rows and then randomize the row order.}

NOTE: These will be used to estimate any shift in the response means between blocks.

\protocolstep{6.2.4}{Concatenate the confirmation runs and the new space-filling runs into a single table and randomize the run order. Use \textbf{Tables \(>\) Concatenate} and then create and sort by a new random column to randomize the run order, as described in Section 2.}

\protocolstep{6.3}{Formulate the new recipes and collect the results.}

\protocolstep{6.4}{Concatenate the new experimental runs and results to the original experiment data table, introducing an experiment-ID column to indicate the source of each result. Use \textbf{Tables \(>\) Concatenate} and select the option to \textbf{Create Source Column}.}

\protocolstep{6.5}{Verify that the column properties for each factor display the combined range over both studies: right-click on the column header for each factor and examine the \textbf{Coding} and \textbf{Mixture} property ranges, if present.}

\protocolstep{6.6}{Begin analysis of the new experiment's results.}

\protocolstep{6.6.1}{Include the experiment-ID column as a term in the model to serve as a blocking factor. Ensure this term does not interact with the study factors. Run the \textbf{Fit Model} dialog script saved to the table in Section 4, select the experiment-ID column and click \textbf{Add} to include it in the list of candidate effects.}

\protocolstep{6.6.2}{Run this \textbf{Fit Model} dialog on the concatenated data table to jointly analyze the results from the new experiment and the initial study. Adhere to earlier instructions to generate updated optimal formulation candidates and graphical summaries.}

\protocolstep{6.6.3}{For validation, independently analyze the results from the new experiment, excluding results from the initial experiment. That is, perform the steps described in Section 4 on the new experimental table.}

\protocolstep{6.6.4}{Ensure that optimal formulations identified by these models align closely with those recognized by the joint analysis.}

\protocolstep{6.6.5}{Review graphical summaries to confirm that both the joint and individual analyses of the new experimental results exhibit similar response surface behaviors (meaning that there is a similar relationship between the response(s) and the factors).}

\protocolstep{6.6.6}{Compare the combined and individual analyses of new results with the initial experiment for consistency. Use similar graph structures for comparison and examine the identified optimal recipes for differences.}

\subsection*{7. Documenting the Study's Final Scientific Conclusions}
\label{protocol:7}

\protocolstep{7.1}{Should the benchmark control change to a newly identified recipe due to the study, log the new setting and specify the design and analysis files that record its origin.}

\protocolstep{7.2}{Maintain all experimental tables and analysis summaries, preferably with date-stamped file names, for future reference.}

\FloatBarrier
\section*{Representative Results}

This approach has been validated across both broadly classified lipid types: MC3-like classical lipids and lipidoids (\textit{e.g.}, C12-200), generally derived from combinatorial chemistry. Compared to a benchmark LNP formulation developed using a One Factor at a Time (OFAT) method, the candidate formulations generated through our workflow can yield large potency improvements. \textbf{Table~\ref{tab:1}} shows one example: across two optimization phases (an initial study and a subsequent follow-up study), mouse liver luciferase expression increased 10,000-fold over the benchmark control on the photon/second scale. In the first phase, we focused on optimizing the lipid ratios while keeping other factors constant. In the follow-up study, we introduced an additional helper lipid type and performed optimization considering both the lipid ratio composition and the helper lipid type. Consequently, the newly introduced helper lipid type was selected to be used with the associated optimized lipid composition. The significant enhancement in potency suggests that these optimized compositions may exhibit superior endosomal escape capabilities\cite{ref27}.

\begin{table}[htbp]
\centering
\includegraphics[width=\linewidth,height=0.67\textheight,keepaspectratio]{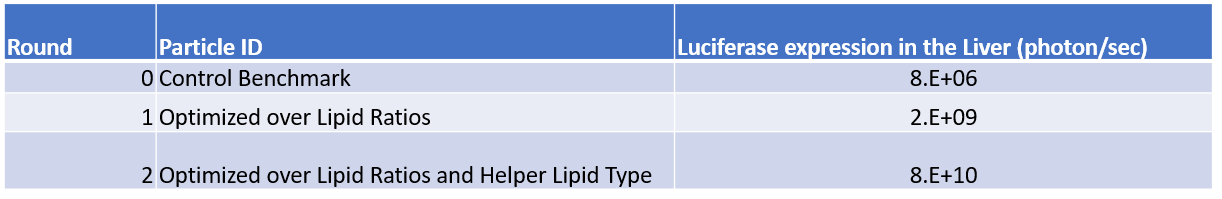}
\caption{Systematic improvement in Luciferase expression through Design of Experiments (DOE) optimization. This table illustrates the significant enhancement in the expression of Luciferase, with up to a 10,000-fold improvement on the photon/second scale, from the initial benchmark to the final ``optimal candidate''.}
\label{tab:1}
\end{table}

\begin{figure}[htbp]
\centering
\includegraphics[width=\linewidth,height=0.67\textheight,keepaspectratio]{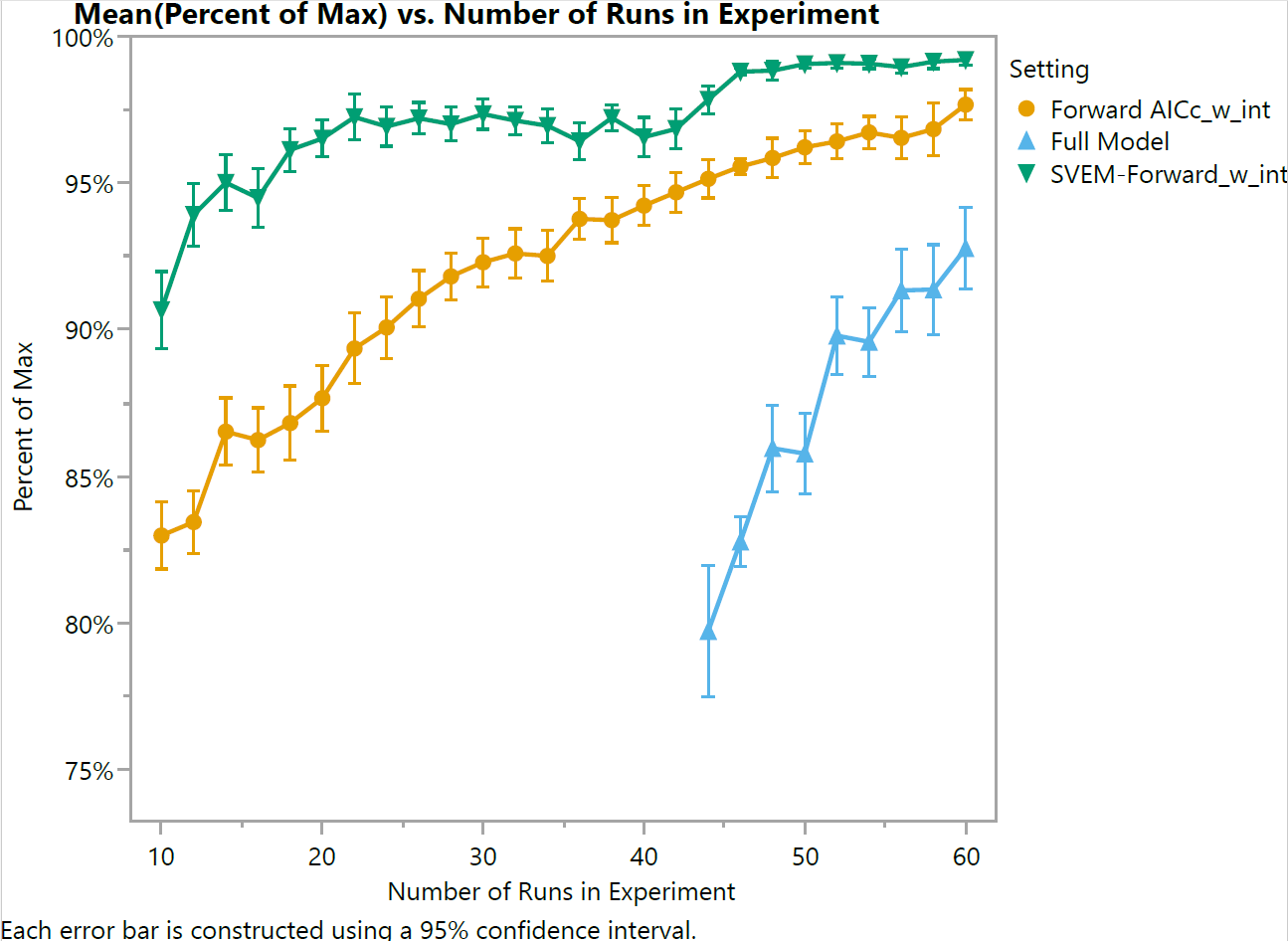}
\caption{Quality of optimal formulation as a function of experimental size and statistical model. The vertical axis represents the percentage of theoretical maximum desirability, and the horizontal axis represents the size of the space-filling design. Each point shows the mean over 150 simulations. The blue line (triangles) represents the full model (without any elimination of statistically insignificant effects), the amber (circles) line represents the traditional AICc-based forward selection model (with an intercept and without forcing mixture main effects), and the green line (upside down triangles) represents the SVEM-based forward selection model (with an intercept and without forcing mixture main effects).}
\label{fig:18}
\end{figure}

Simulations can be used to show the expected quality of the optimal candidate produced by this procedure. Within the framework of the example experiment used in the protocol, we can repeat the simulation many times for different run sizes and evaluate the results according to the simulated process-generating function. A JMP script for this purpose is provided in \textbf{Supplementary File~4}. Specifically, we generate a space-filling design and populate the response columns with values from our generator functions, plus noise representing analytical and process variation. We fit these simulated responses with different analysis techniques (including SVEM Forward Selection) to produce a corresponding candidate optimal recipe. The candidates from each analysis method are then compared to the value of the true optimum from the generating functions. \textbf{Figure~\ref{fig:18}} illustrates the average percent of the maximal theoretical desirability achieved by each of three analysis methods using space-filling designs of size given on the horizontal axis. The full model, which includes all candidate effects and does not reduce the model based on the statistical significance of those effects, performs the worst. Much of the additional work that traditionally goes into fitting regression models for mixture-process experiments involves modifications (removing the intercept, forcing the mixture main effects, precluding the use of pure quadratic mixture effects, etc.) that are required to fit this full model\cite{ref09}, and from this perspective those procedures are unnecessary\cite{ref12}. Furthermore, this model cannot be fit until the design size reaches the number of effects in the model. At smaller experimental sizes, we can fit the traditional forward selection method, which outperforms the full model with respect to average performance of the optimal candidate formulation for each fixed experimental size. Likewise, the SVEM modification to this forward selection approach further improves the performance of the optimal candidates. This plot reveals that using SVEM-Forward Selection\cite{ref12,ref13} to analyze a 24-run space-filling experiment achieves the same average quality typically requiring 50 runs when analyzed with a traditional forward selection (targeting minimum AICc) model. Although actual performance will vary from process to process, this simulation -- along with published results on SVEM\cite{ref12,ref13,ref16,ref17,ref28} -- demonstrates the potential of this modeling procedure for formulation optimization.

\FloatBarrier
\section*{Supplementary Files}

\textbf{Supplementary File 1: 04APR2023 Summary.docx} -- This document provides a record of the study including its purpose, the responses assessed, the factors considered, and the total number of runs executed.

\textbf{Supplementary File 2: 23\_run\_simulated\_experiment.jmp} -- A JMP file with the simulated experiment and its results. This file also includes attached analysis scripts compatible with JMP 17 Pro.

\textbf{Supplementary File 3: 23\_run\_simulated\_experiment.xlsx} -- An Excel file that includes the simulated experiment and its results, suitable for readers who may not have access to JMP.

\textbf{Supplementary File 4: mixture simulation 20DEC22.jsl} -- This is a JMP 17 Pro script used to simulate LNP formulation experiments and evaluate the performance of different analysis methods. The script compares several analysis methods, including SVEM-Forward Selection with and without an intercept; the workflow presented here uses the version with an intercept.

\FloatBarrier
\section*{Discussion}

Modern software for the design and analysis of mixture-process experiments makes it possible for scientists to improve their lipid nanoparticle formulations in a structured workflow that avoids inefficient OFAT experimentation. The recently developed SVEM modeling approach eliminates many of the arcane regression modifications and model reduction strategies that may have previously distracted scientists with extraneous statistical considerations. Once the results are collected, the SVEM analysis framework offers an approach that is both easier to implement and tends to produce better models than traditional modeling approaches\cite{ref13}. Furthermore, the graphical analyses that are based on the prediction formulas for each response are easily interpretable by scientists, giving a clear summary of the marginal behavior of the response over individual factors as well as small groups of factors without requiring the interpretation of highly correlated parameter estimates from a regression model. This allows scientists to focus on assessing practical significance across study factors after SVEM has selected or shrunk candidate effects for prediction.

The workflow has been used in practice to systematically vary lipid composition and formulation parameters such as N:P ratio, flow rate, and mixing ratio for optimization and to select the best helper lipid types, ionizable lipid types, and buffer types. The goals across these examples usually include maximizing \textit{in vivo} or \textit{in vitro} potency and encapsulating varying payloads like mRNA or DNA for relevant \textit{in vivo} targets such as liver cells, or sometimes across multiple cell types in the case of \textit{in vitro} applications. For specific applications, we may need to balance biophysical properties such as size, PDI, zeta potential, and percent encapsulation while examining \textit{in vivo} potency. Additionally, the goal is to find a potent, yet well-tolerated formulation and so we may include responses such as change in body weight, cytokine response, or elicitation of liver enzymes such as AST/ALT in the analysis. Patterns have emerged from numerous LNP experiments. Notably, alterations in the molar ratio of the ionizable lipid and the N:P ratio seem to significantly impact RNA encapsulation. Moreover, changes in the PEG molar ratio appear to affect particle stability, as indicated by influences on size and PDI. In general, an excess of PEG in the LNP core tends to have a detrimental effect on potency in mice.

Performance improvements are especially noticeable when more than one response is targeted: even if the benchmark already performs well with respect to the primary response (\textit{e.g.}, potency), joint optimization typically maintains or improves the behavior with respect to the primary response while simultaneously improving behavior with respect to other responses (minimizing PDI, size, or bodyweight loss). We validate the authenticity of these improvements with confirmation runs, wherein we prepare and directly compare the benchmark formulation (possibly with a replicate) and new candidate formulations.

The design phase of this workflow has several critical steps. First, ensure that the factors and their ranges are correctly entered into the space-filling design platform. Second, use graphics and subject matter knowledge to confirm the feasibility of each resulting formulation before initiating the experiment. Finally, execute the experiment following the randomized order specified by the design table. Adhering to this sequence helps prevent unmeasured covariates---such as the order of formulation production or ambient temperature---from confounding the factors under study. Space-filling designs are easier to construct, with less potential for user error, than optimal mixture-process designs, which require extra decisions during setup that may frustrate inexperienced users and discourage them from using designed experiments. Nevertheless, after working through this protocol, scientists may benefit from additional reading on how optimal designs could potentially replace space-filling designs in the protocol, such as described in Chapter 6 of Goos and Jones (2011)\cite{ref29}. Especially for follow-up studies that ``zoom in'' on an optimal region -- where there is less concern about failures along the mixture boundaries -- D-optimal designs can be more efficient than space-filling designs.

Likewise, the analysis phase of this workflow has several critical steps. First, ensure that the model specifies an appropriate set of candidate effects, including interactions, rather than only the main (first-order) effects of the factors. Second, employ SVEM Forward Selection as the modeling framework. Third, disable the default \textit{No Intercept} option and avoid forcing mixture main effects. Finally, correctly set the desirability functions for the responses before initiating optimization. When SVEM is unavailable in the software being used, the best approach is to use traditional forward selection (targeting minimum AICc) for the regression problem\cite{ref12}. The protocol mentions that it is also possible to use SVEM Lasso: on average, this approach gives similar results to SVEM Forward Selection, though for particular datasets the two approaches may produce slightly different optimal formulations that could be compared with confirmation runs\cite{ref12}. However, SVEM Lasso will give inferior modeling results if the user makes the easy mistake of forgetting to disable the default \textit{No Intercept} option\cite{ref12}: for this reason, we have used SVEM Forward Selection as the default method, since it is more robust to this option.

The primary limitation of this method is that there will be occasional studies with greater complexity that will benefit from the help of a statistician for design and analysis. Situations where the run budget is more limited than usual (below the minimum heuristic), the responses are binary, there are a large number of categorical factors or levels of a single categorical factor, where a research goal is to consider eliminating one or more mixture factors from the recipe, or where there are additional constraints on the factor space may be approached differently by a statistician, such as by using optimal or hybrid\cite{ref12,ref30} designs or by adding additional structure to the design. Specifically, a hybrid design could be formed by creating a space-filling design with most of the budgeted runs and then ``augmenting'' the design with the remaining runs (usually 2-4) using a D-optimal criterion. Another hybrid approach is to generate a space-filling design over the mixture (lipid) and continuous (process) factors, and then afterward to add any categorical factors using an ``optimal'' allocation of factor levels. Nevertheless, the simplified space-filling design approach taken in the protocol has been developed over the past few years in the process of running dozens of LNP formulation optimization experiments, and we believe it offers a robust approach that will work successfully in most cases while also giving scientists confidence in their ability to utilize designed experiments.

\FloatBarrier
\section*{Disclosures}

The experimental design strategy underpinning this workflow has been employed in two patent applications in which one of the authors is an inventor\cite{ref25,ref26}. Additionally, Adsurgo, LLC is a certified JMP Partner. However, the development and publication of this paper were undertaken without any form of financial incentive, encouragement, or other inducements from JMP.

\FloatBarrier

\end{document}